\documentclass[aps,prb,twocolumn,superscriptaddress]{revtex4-2}

\usepackage[utf8]{inputenc}

\usepackage{mathtools}
\usepackage{amsfonts}
\usepackage{mathrsfs}
\usepackage{physics}
\usepackage{slashed}
\usepackage{tensor}
\usepackage{dsfont}
\usepackage{bbm}

\usepackage{graphicx}
\usepackage{color, float}
\usepackage{array}
\usepackage[abs]{overpic}

\usepackage{tikz}
\usetikzlibrary{decorations.pathmorphing}

\usepackage{placeins}

\usepackage{makecell}

\usepackage{xspace}
\usepackage{siunitx}
\usepackage{xfrac}
\usepackage{hyperref}
\usepackage[nameinlink]{cleveref}
\usepackage{appendix}

\usepackage{adjustbox}

\usepackage{xifthen}
\usepackage{xcolor}
\hypersetup{
	colorlinks,
	linkcolor={red!75!black},
	citecolor={blue!75!black},
	urlcolor={blue!75!black}
}

\newcommand{\gettitle}{One-dimensional long-range Ising model: two (almost) equivalent approximations}

\newcommand{\getZuerichAffiliation}{\affiliation{Institut f{\"u}r Theoretische Physik, ETH Z{\"u}rich, Wolfgang-Pauli-Str. 27, 8093 Z{\"u}rich, Switzerland}}

\newcommand{\getENSAffiliation}{\affiliation{Laboratoire de Physique de l'\'Ecole Normale Sup\'erieure, CNRS, ENS $\&$ PSL University, Sorbonne Universit\'e, Universit\'e Paris Cité, 75005 Paris, France}}

\newcommand{\getTriesteUniAffiliation}{\affiliation{Dipartimento di Fisica, Universit{\`a} di Trieste, Strada Costiera 11, I-34151 Trieste, Italy}}

\newcommand{\getSissaAffiliation}{\affiliation{SISSA and INFN Sezione di Trieste, Via Bonomea 265, I-34136 Trieste, Italy}}

\newcommand{\getDemocritosAffiliation}{\affiliation{CNR-IOM DEMOCRITOS Simulation Center, Via Bonomea 265, I-34136 Trieste, Italy}}

\newcommand{\getCNR}{\affiliation{CNR-INO, Area Science Park, Basovizza, 34149 Trieste, Italy}}

\hypersetup{
	pdftitle={\gettitle},
	pdfauthor={Pagni},
	pdfkeywords=
	{hierarchical model} {renomalization group} ,
	bookmarksopen=true,
	bookmarksopenlevel=2,
	bookmarksnumbered=true
}

\begin{document}
	
\title{\gettitle}

\author{Valerio Pagni}\getZuerichAffiliation
\author{Guido Giachetti}\getENSAffiliation
\author{Andrea Trombettoni}\getTriesteUniAffiliation \getSissaAffiliation \getDemocritosAffiliation
\author{Nicolò Defenu}\getZuerichAffiliation \getCNR

\begin{abstract}
	We investigate the critical behavior of the one-dimensional Ising model with long-range interactions using the functional renormalization group in the local potential approximation (LPA), and compare our findings with Dyson’s hierarchical model (DHM). While the DHM lacks translational invariance, it admits a field-theoretical description closely resembling the LPA, up to minor but nontrivial differences. After reviewing the real-space renormalization group approach to the DHM, we demonstrate a remarkable agreement in the critical exponent $\nu$ between the two methods across the entire range of power-law decays $1/2 < \sigma < 1$. We further benchmark our results against Monte Carlo simulations and analytical expansions near the upper boundary of the nontrivial regime, $\sigma \lesssim 1$.
\end{abstract}

\maketitle

\section{Introduction}\label{sec1}

Long-range interacting systems have emerged as a central theme in modern many-body physics, exhibiting a rich variety of collective phenomena that challenge and extend the standard paradigms established for short-range systems~\cite{defenu2023long}. These include the breakdown of locality, violations of conventional scaling laws, and the emergence of novel universality classes. Such systems naturally arise in numerous physical contexts, including trapped ion chains~\cite{britton2012engineered}, dipolar gases~\cite{lahaye2009physics}, Rydberg atom arrays~\cite{browaeys2020many}, and cavity QED setups~\cite{ritsch2013cold}. Moreover, recent progress in quantum information and nonequilibrium dynamics has further highlighted the role of long-range interactions in phenomena such as dynamical phase transitions, thermalization and equilibration, and information spreading beyond the light-cone limit, see Refs.~\cite{defenu2023long,defenu2024out} for a review. These developments have renewed interest in understanding the critical behavior of long-range models, both at and out of equilibrium, and in developing robust theoretical tools to describe them.

Among the various models used to investigate long-range critical phenomena, a particularly influential and extensively studied example is the one-dimensional long-range Ising (1D LRI) model, defined by the Hamiltonian $H = -\frac{1}{2} \sum_{i \neq j} J_{ij} s_i s_j$ with spin variables $s_i=\pm 1$
and couplings decaying algebraically as \begin{equation}\label{eq:LR_Hamiltonian}
	J_{ij} = \frac{J}{|i - j|^{1 + \sigma}}, \qquad (J>0)
\end{equation}  
where \(\sigma > 0\) ensures a well-defined thermodynamics~\cite{campa2009statistical}. Unlike its short-range counterpart in $d = 1$, the 1D LRI model exhibits a finite-temperature phase transition for certain values of $\sigma$. In fact, the long-range nature of the couplings suppresses the strong fluctuations typical of low-dimensional systems, allowing for spontaneous symmetry breaking. This was first rigorously established by Dyson in Ref.~\cite{dyson1969existence} for all $\sigma \in (0,1)$, through the introduction of Dyson's hierarchical model (DHM) as a tractable lower bound. For $\sigma >1$ there is no transition~\cite{ruelle1968statistical}. The marginal case $\sigma = 1$, corresponding to interactions decaying as $1/r^2$, remained unresolved in Dyson’s original analysis. It was later shown by Fröhlich and Spencer~\cite{frohlich1982phase} that this case features a Berezinskii–Kosterlitz–Thouless (BKT)-like transition, characterized by an essential singularity in the free energy. Notably, however, this transition includes a discontinuity in the magnetization -- anticipated by Thouless in Ref.~\cite{thouless1969long}, see also~\cite{dyson1971ising} -- which distinguishes it from the standard BKT transition in the two-dimensional XY model. The strong interest in the Ising model with $1/r^2$ interactions was the result of its equivalence to the Kondo problem, established by Anderson, Yuval and Hamann in a series of papers~\cite{yuval1970exact,anderson1970exact,anderson1971some}, which also contain an early version of the renormalization group (RG) later developed by Wilson~\cite{wilson1974renormalization}. Their methods were extended by Cardy~\cite{cardy1981one} to encompass general discrete spin chains with $1/r^2$ interactions, including Potts and Ashkin–Teller models.

%However, 
The main interest of this work is the critical behavior of the 1D LRI with $0<\sigma < 1$. This is related to the quantum criticality of the spin-boson model~\cite{leggett1987dynamics} -- via an imaginary time path integral approach analogous to~\cite{yuval1970exact,anderson1970exact,anderson1971some} -- where the power-law exponent of the spectral function of the bosonic bath corresponds to $\sigma$ in the classical 1D LRI model. Despite the absence of an exact solution for the partition function, the critical physics of the 1D LRI with power law exponent $\sigma$ has been investigated through various powerful techniques, 
such as Monte Carlo (MC) simulations~\cite{luijten1997classical,uzelac2001critical,tomita2008monte,angelini2014relations}, RG approaches~\cite{kosterlitz1976phase,cannas1995one}, and, recently, conformal field theory (CFT) methods~\cite{benedetti2025one,benedetti2025strong}. 

In this work, we adopt the RG framework, which has played a central role in the study of long-range critical systems~\cite{fisher1972critical,sak1973recursion}. Our goal is twofold. First, to compute the critical exponent $\nu$ within a functional renormalization group (FRG) approach~\cite{wetterich1993exact,tetradis1994critical,berges2000nonperturbative,dupuis2021the} -- notice, indeed, that one expects that the critical exponent $\eta$ has a trivial dependence on $\sigma$~\cite{sak1973recursion} and therefore the main critical exponents are determined once the non-trivial dependence of $\nu$ has been worked out. Second, to explore the connection between the FRG analysis of the 1D LRI model and the real-space RG treatment of Dyson's hierarchical model, which is a well-developed tool~\cite{kim1977critical,meurice2007nonlinear}, even from a mathematically rigorous point of view~\cite{bleher1973investigation,bleher1975critical,bleher2010critical}. Although the DHM is not translationally invariant and is often regarded as a toy model, we argue that it admits an effective description that closely mirrors the local potential approximation (LPA) of the FRG.

The structure of the paper is as follows. After a brief reminder in Sec.~\ref{sec:added} of results on critical properties of long-range systems useful for the rest of the paper, in Sec.~\ref{sec2} we introduce the FRG formalism and present a non-perturbative estimate of the critical exponent $\nu=\nu(\sigma)$, which, to our knowledge, has not been explicitly computed for the 1D LRI model in this context. In Sec.~\ref{sec3}, we argue for a field-theoretical formulation of the DHM and show how it is closely related to the FRG in the LPA. We then review the calculation of $\nu$ in the DHM by means of real-space RG and compare it with the FRG results. In Sec.~\ref{sec4}, we benchmark our findings against Monte Carlo simulations and analytical expansions near the $\sigma \to 1$ and $\sigma \to 1/2$ limits. Our conclusions and outlook are presented in Sec.~\ref{sec5}.

\section{Reminder on critical properties of long-range systems}
\label{sec:added}

In this %e remaining part of this 
section, we %provide 
briefly review results on the characterization of critical behavior 
of long-range systems~\cite{campa2009statistical,defenu2023long} and %briefly 
discuss the case with spatial dimension $d=1$, which presents some peculiarities.

Focusing on the regime $\sigma > 0$, where interactions decay more rapidly than $1/r^d$, we distinguish between three scenarios. In the interval $\sigma \in (0, \sigma_{\rm mf})$ the critical behavior is essentially mean-field and governed by the Gaussian fixed point, very similar to how mean-field theory provides the correct exponents above the upper critical dimension in short-range systems. Distinctive long-range features emerge for $\sigma \in (\sigma_{\rm mf}, \sigma_*)$, where the critical exponents depend continuously on the range parameter $\sigma$. In this intermediate regime -- the focus of this work -- the critical fixed points are interacting ones, meaning that the role of fluctuations becomes crucial. Finally, for $\sigma \in (\sigma_*,\infty)$ the critical behavior crosses over to the short-range universality class, as the long-range interaction becomes irrelevant. While $\sigma_{\rm mf} = d/2$ for classical LR $O(n)$ models~\cite{defenu2023long,campa2009statistical}, the exact value of $\sigma_*$ has long remained controversial in $d>1$\,\cite{defenu2020criticality}. Today, the most widely accepted scenario, proposed by Sak~\cite{sak1973recursion}, is that $\sigma_* = 2-\eta_{\rm SR}$, where $\eta_{\rm SR}$ is the anomalous dimension of the model with short-range interactions, while for the long-range model $\eta(\sigma) = 2-\sigma$\,\cite{angelini2014relations,defenu2015fixed,behan2017long}.
Critical properties of $d$-dimensional long-range systems have been studied by a variety of methods, including Monte Carlo~\cite{luijten1997interaction,fukui2009order,angelini2014relations,horita2017upper}, RG approaches~\cite{fisher1972critical,sak1973recursion,kosterlitz1976phase,cannas1995one,defenu2015fixed} and conformal bootstrap~\cite{behan2017long,behan2017scaling,Behan_2019,Behan2024}; see further references in~\cite{campa2009statistical,defenu2020criticality,defenu2023long}. Peculiar behavior may arise in two-dimensional $O(2)$ symmetric models\,\cite{Giachetti2021,Giachetti2022prb,Giachetti2021EPL,giachetti2023villain,tianning2025two}, which is yet the subject of active research.

In $d=1$, the values $0< \sigma<1/2$ identify the region where critical exponents are exactly given by mean-field theory, e.g.~$\nu(\sigma) = 1/\sigma$ for the 1D LRI model (notice that this region for long-range bond-percolation in $d=1$ is $0< \sigma<1/3$~\cite{gori2017one}). On the other hand, the %non-classical 
non-mean-field region extends only across $1/2 < \sigma < \sigma_* = 1$, as anticipated by the foregoing summary of results about the 1D LRI model. In fact, this agrees with~\cite{sak1973recursion}, as formally we have $\eta_{\rm SR}(d=1) = 1$, due to the fact that in one dimension the short-range model has only a zero-temperature transition, where the two-point correlator decays as a power-law with vanishing exponent, i.e.~it is constant.
% ($r^{-\frac{d-2+\eta_{\rm SR}}{2}} = r^{-\frac{-1+\eta_{\rm SR}}{2}} = \text{const.}$)

The lack of a finite-temperature transition in the short-range Ising chain carries interesting consequences. In the absence of a short-range universality class, the usual picture consisting in the crossover from a line of long-range fixed points to the short-range regime, see e.g.~\cite{behan2017long}, becomes more involved. The crucial observation is that in the region around $\sigma_* = 1$ the most suitable degrees of freedom are not the original spins, but the domain walls, or kinks, introduced in~\cite{yuval1970exact,anderson1970exact,anderson1971some} for $\sigma = 1$, and used in the RG analysis of~\cite{kosterlitz1976phase} for $\sigma \lesssim 1$, which considers a diluted gas of alternating kinks. As an extension of this approach, and based on the infrared duality of~\cite{behan2017long} for $d>1$, Refs.~\cite{benedetti2025one,benedetti2025strong} provide a weakly-coupled field theory for the 1D LRI model below $\sigma=1$. Contrary to higher dimensional LRI models, their theory is written in terms of a compact free field with negative scaling dimension, perturbed by vertex operators with the alternating-kink constraint, which is enforced via an algebra of Pauli matrices. Thus, we expect that the peculiar character of the one-dimensional problem close to the short-range limit is going to also affect the following analyses -- based on the RG.

\section{Local potential approximation}\label{sec2}

\subsection{Functional renormalization group in the LPA}

In the first part of this article, we study the critical physics of the 1D LRI model in the %non-classical 
regime $\sigma \in (1/2,1)$ by means of a functional renormalization group (FRG) approach based on the effective average action. The latter is a modification of the generating functional $\Gamma[\phi]$ of one-particle irreducible vertex functions (see e.g.~\cite{amit2005field}), or the Legendre transform of the generating functional of connected correlation functions. Such modification involves considering an external momentum scale $k \geq 0$, and adding a scale-dependent mass term to the bare action $S_{\rm bare}[\phi]$ describing the field theory, so that the effective action reads $\Gamma_k[\phi]$, as it carries a $k$-dependence. The mass term dependent on $k$ is included to regularize infrared modes with momentum scale $q \ll k$. To achieve that, a regulator function $R_k(q)$ with appropriate features has to be specified. In particular, we want to choose $R_k$ so that at the microscopic scale $k = \Lambda$ one retrieves the action $S_{\rm bare}[\phi] = \Gamma_{k=\Lambda}[\phi]$, while at the largest scales $k\to0$ the full generating functional is approached, that is $\Gamma_{k \to 0}[\phi] \to \Gamma[\phi]$.

Within this setup it is possible to write %a powerful identity, 
the functional and non-perturbative RG equation
\begin{align}\label{eq: wetterich}
	\partial_t \Gamma_k[\phi] = \frac{1}{2} \int_{{q}} \tr \left[{G}_k({q}) \, \partial_t {R}_k({q}) \right],
\end{align}
proposed by Wetterich and Morris~\cite{wetterich1993exact,morris1994exact}, where $\int_{q} = \int  dq /(2\pi)$ and $t=\log(k/\Lambda)$. Denoting the second functional derivative (Hessian) of the action by $\Gamma_k^{(2)}$, the field-dependent propagator $G_k = G_k[\phi]$ reads
\begin{equation}\label{eq: full propagator}
	{G}_k[{\phi}] = \left( \Gamma_k^{(2)}[{\phi}] + {R}_k \right)^{-1}.
\end{equation}
More details are found in~\cite{tetradis1994critical,berges2000nonperturbative,dupuis2021the} and references therein.

While there is no general solution to the equation~\eqref{eq: wetterich}, it is possible to consider a truncated form of the action, so that it becomes possible to project the RG onto functions -- rather than functionals -- and obtain simpler differential equations. A standard ansatz chosen for $\Gamma_k$ is the \textit{local potential approximation} (LPA), which for the long-range Ising model in one spatial dimension is
\begin{equation}\label{eq: LPA ansatz}
	\Gamma_k[\phi] = \int_{-\infty}^{\infty} dx \, \left\{ \phi(x) (-\Delta)^{\sigma/2} \phi(x) + V_k(\phi(x)) \right\},
\end{equation}
where $V_k(\phi)$ is a local function of the field called effective potential, and $(-\Delta)^{\sigma/2}$ is the fractional Laplacian~\cite{lischke2020fractional}. For an infinite homogeneous chain we equivalently use the momentum-space variant of the gradient term:
\begin{equation}\label{eq: gradient term}
	\int_{-\infty}^{\infty} \frac{dq}{2\pi} \, q^{\sigma} \phi(q) \phi(-q) = \int_{-\infty}^{\infty} dx \, \phi(x) (-\Delta)^{\sigma/2}  \phi(x),
\end{equation}
whence it becomes clear that long-range interactions are described by a continuous field theory with non-analytic momentum dependence. Note that $q^\sigma \equiv |q|^{\sigma}$.

It is crucial to note that the $k$-dependence of the LPA ansatz~\eqref{eq: LPA ansatz} lies entirely in the effective potential $V_k(\phi)$. In other words, the gradient term~\eqref{eq: gradient term} is \textit{not} subject to any renormalization. The assumption that the theory is given by~\eqref{eq: LPA ansatz} is certainly an approximation, so we cannot expect it to capture exactly the critical behavior of the model. In fact, the LPA is the leading order of the so-called derivative expansion, which would in principle enable a systematic improvement of numerical results by the inclusion of higher-order derivative interactions.

For long-range theories, however, the next to leading order improvement of the LPA does not come from a renormalization of the gradient term as in short-range models. In fact, it has been shown~\cite{defenu2015fixed} that no anomalous scaling of the non-analytic term is generated by the FRG if a (field-independent) wavefunction renormalization $Z_k$ is introduced in front of the $q^\sigma$ term. This also agrees with~\cite{honkonen1989crossover}, where a similar result is shown for Wilson's RG to all orders of perturbation theory. On the other hand, the short-range analytic term of the form $q^2 \phi(q) \phi(-q)$ is always generated under RG~\cite{honkonen1989crossover} together with other, sub-leading, non-analytic terms\,\cite{balog2014critical}. We are neglecting this term in the present analysis. In dimension $d>1$ its influence becomes noticeable only very close to $\sigma = \sigma_*$~\cite{solfanelli2024universality}.

\subsection{RG flow and fixed points} %of the potential}

As previously discussed, we can now use the general RG equation to deduce the flow of the effective potential. Evaluating Eq.\,\eqref{eq: wetterich} at a constant field configuration, the gradient term disappears, leading to
\begin{equation}\label{eq: flow of potential (general)}
	\partial_t V_k(\phi) = \frac{1}{2} \int_q \frac{\partial_t R_k (q)}{q^{\sigma} + R_k(q) + V_k''(\phi)}.
\end{equation}
Let us now consider the regulator~\cite{defenu2015fixed}
\begin{equation}\label{eq: Litim regulator}
	R_k(q) = (k^\sigma-q^\sigma) \, \theta (k^\sigma-q^\sigma)
\end{equation}
generalizing the standard Litim cutoff used for short-range systems. We employ this regulator because it satisfies an optimization criterion for the LPA~\cite{litim2000optimization,litim2001optimized,nandori2014spontaneous}, thereby minimizing systematic errors in the computation of critical exponents~\cite{dupuis2021the}. For example, the critical exponents found with the Litim regulator are more accurate \cite{defenu2018scaling} than those obtained with a power-law cutoff \cite{morris1994derivative,morris1995renormalization}. Most importantly for our purposes, this optimized choice is structurally closest to the recursion relations of the DHM, allowing for the direct comparison performed in Sec.~\ref{sec: DHM vs LPA}. The investigation of different regulator choices and their impact on the specific numerical values of critical exponents is deferred to future studies. Focusing here on the optimized Litim cutoff, the flow equation Eq.~\eqref{eq: flow of potential (general)} reads
\begin{equation}\label{eq: flow of potential (Litim)}
	\partial_t V_k(\phi) = \frac{\sigma}{2\pi} \frac{k^{1+\sigma}}{k^{\sigma} + V_k''(\phi)}.
\end{equation}
The %physics 
properties of criticality are captured by certain fixed points of the RG flow. At such fixed points we expect scale-invariant solutions of Eq.~\eqref{eq: flow of potential (Litim)}. We therefore turn~\eqref{eq: flow of potential (Litim)} into the following autonomous differential equation (meaning that the scale $k$ does not appear explicitly in the equation):
\begin{equation}\label{eq: autonomous RG equation}
	\partial_t \tilde{V}_k(\varphi) =  - \tilde{V}_k(\varphi)  + \frac{1-\sigma}{2} \varphi \tilde{V}_k'(\varphi) + \frac{\sigma}{2\pi} \frac{1}{1 + \tilde{V}_k''(\varphi)},
\end{equation}
where we have used the dimensionless variables
\begin{align}\label{dimensionless variables}
	\tilde{x} = k x,
	\qquad
	\varphi(\tilde{x}) = k^{\frac{\sigma-1}{2}} \phi(x),
	\qquad
	\tilde{V}_k = k^{-1} V_k.
\end{align}
Setting the left-hand side of~\eqref{eq: autonomous RG equation} to zero yields the fixed-point equation. This is now an ordinary differential equation whose solutions $\tilde{V}_*(\varphi)$ can be studied by means of a shooting method known as spikeplot~\cite{morris1994truncations,codello2012scaling,defenu2015fixed} or via pseudospectral approaches~\cite{borchardt2015global,ihssen2025nonperturbative}. Both the Gaussian and the interacting (Wilson-Fisher) fixed points are found without the need of any Taylor expansion of the effective potential.

\subsection{Linearization around the fixed point}

Once the scaling solution $\tilde{V}_*(\varphi)$ has been obtained, our objective is to study the neighborhood of the interacting fixed point in order to obtain the critical exponent in the regime $1/2 < \sigma < 1$. One of the exponents, namely the anomalous dimension $\eta = 2-\sigma$, is already determined thanks to the analysis of~\cite{defenu2015fixed}, which holds in our case as well. We are therefore interested in determining $\nu$, so that the remaining exponents can be retrieved by the use of scaling laws, such as $\gamma = \nu(2-\eta) = \nu \sigma$.

\begin{figure}
	\centering
	\includegraphics[width=1.0\linewidth]{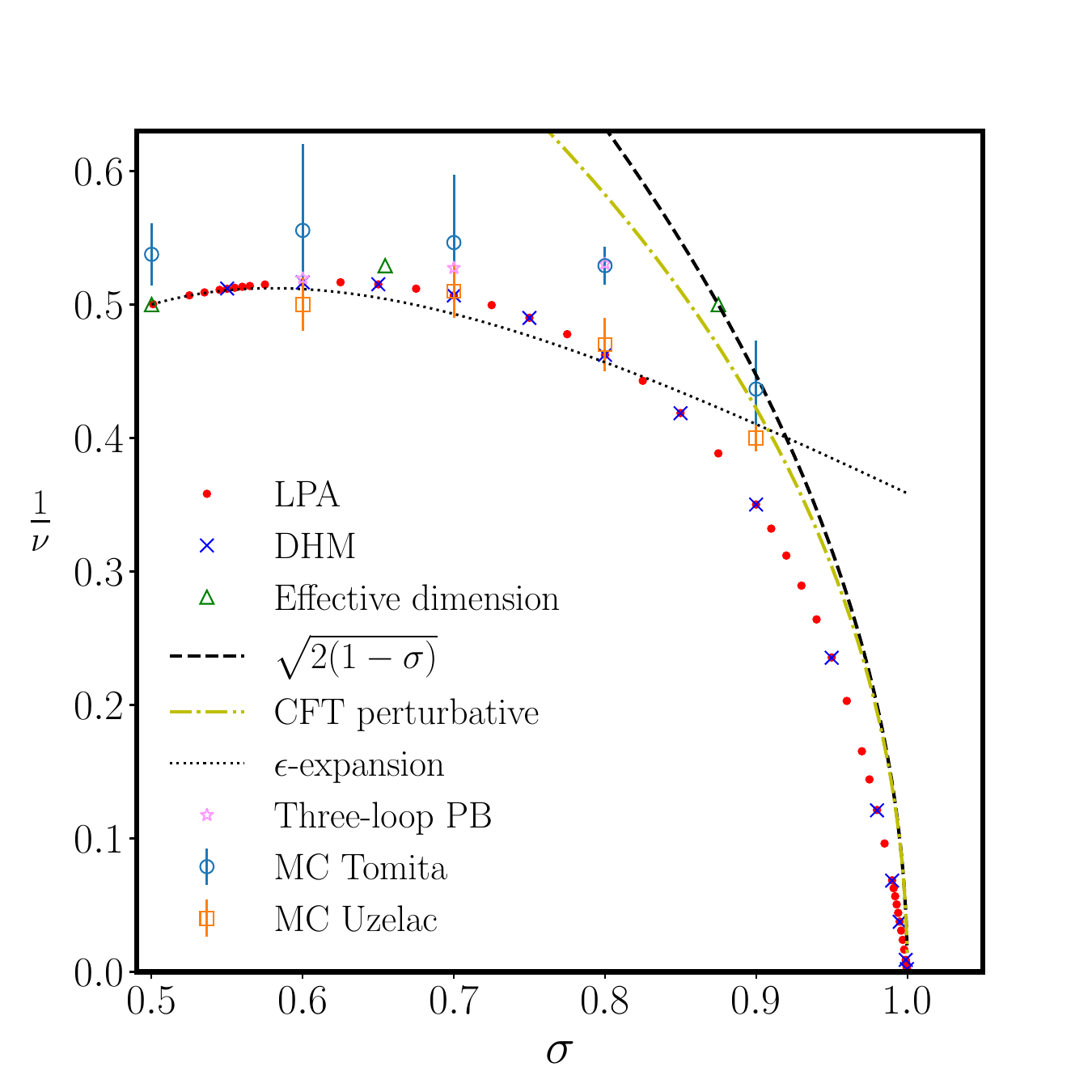}
	\caption{Inverse of the critical exponent $\nu$ of the 1D LRI model in the region $1/2 < \sigma < 1$. The red dots are the results coming from a non-perturbative FRG calculation at the LPA level, described in Sec.~\ref{sec2}. The blue crosses represent the estimates obtained from the DHM via a real-space RG, see Sec.~\ref{sec3}. The green triangles at $\sigma = 0.5$, $\sigma \approx 0.65$ and $\sigma = 0.875$ are obtained via the effective dimension approach, Eq.~\eqref{eq:effdim}. The dotted line is the two-loop $\epsilon$ expansion~\eqref{eq: epsilon expansion} of~\cite{fisher1972critical}, while the pink stars are the three-loop Padé-Borel resummed exponents of~\cite{benedetti2025addendum}. The black dashed and olive dash-dotted lines are the results coming from expansions around $\sigma = 1$, respectively from~\cite{kosterlitz1976phase} and~\cite{benedetti2025one}. More details about these are provided in Sec.~\ref{sec4}. A detailed comparison of LPA and DHM results is presented in Fig.~\ref{fig:differencelpahm}. Finally, Monte Carlo points are taken from Refs.\,\cite{tomita2008monte} and \,\cite{uzelac2001critical}.}
	\label{fig:exponentnu}
\end{figure}

The linearization around the fixed point $\tilde{V}_*(\varphi)$ leads to the determination of RG eigenvalues~\cite{wilson1974renormalization}, the largest of which (typically known as thermal eigenvalue) corresponds to the inverse of $\nu$: $y_{\text{max}} \equiv y_t = \nu^{-1}$. The FRG approach allows us to obtain the spectrum of eigenvalues functionally, without resorting to truncations of Taylor series. We write the ansatz
\begin{equation}\label{eq: linearization FRG}
	\tilde{V}_k(\varphi) = \tilde{V}_*(\varphi) + \varepsilon e^{-y_n t} u_n(\varphi),
\end{equation}
where $\varepsilon \ll 1$, and $y_n$ is the $n$-th eigenvalue with eigenfunction $u_n$. Substituting~\eqref{eq: linearization FRG} into~\eqref{eq: autonomous RG equation} while keeping only linear terms in $\varepsilon$ leads to
\begin{equation}\label{eq: eigenperturbation equation}
	(y_n-1) u_n(\varphi) + \frac{1-\sigma}{2} \varphi u_n'(\varphi) = \frac{\sigma}{2\pi} \frac{u_n''(\varphi)}{[1+\tilde{V}''_*(\varphi)]^2}.
\end{equation}
Solving this linear differential equation yields the values of $y_{t} = \nu^{-1}$ shown as red dots in Fig.~\ref{fig:exponentnu} for the whole range $\sigma \in (0.5,1)$. In Fig.~\ref{fig:exponentnu} we also report other estimates of $\nu^{-1}$ that will be discussed in the forthcoming sections.

Close to the mean-field limit $\sigma = 1/2$, where the relation $\nu = 1/\sigma$ holds~\cite{fisher1972critical}, we obtain indeed values close to $\nu^{-1} = 0.5$. On the other hand, we note that the exponent $\nu$ diverges in the proximity of $\sigma = 1$, so that $\nu^{-1} \to 0$ for $\sigma \to 1^{-}$, as it should. This is analogous to the short-range XY model, in which case the approach to the BKT transition as $d\to 2$ from above is accompanied by a similar behavior of $\nu$. We observe that near $\sigma = 1$ the LPA results are well fitted by 
$\nu^{-1} \approx {\cal C} (1-\sigma)^\gamma$ with ${\cal C} \approx 4.0$ and $\gamma \approx 0.9$, to be compared with the result 
$\nu^{-1} =\sqrt{2(1-\sigma)}$ of~\cite{kosterlitz1976phase,benedetti2025one}.

\section{Hierarchical model}\label{sec3}

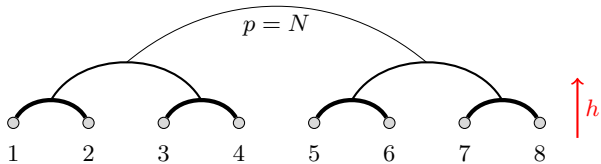
\begin{figure}
	\centering
	\begin{tikzpicture}[scale=1, every node/.style={circle, draw, fill=gray!30, inner sep=1.5pt}]
		
		% Define spacing
		\def\xsep{1}
		\def\ybase{0}
		\def\yarcone{0.75}
		\def\yarctwo{1.8}
		
		% Draw sites
		\foreach \i in {1,...,8} {
			\node (s\i) at ({\i*\xsep}, \ybase) {};
			\node[draw=none, fill=none] at ({\i*\xsep}, \ybase - 0.4) {\i};
		}
		
		% Level p=1:
		\foreach \i/\j in {1/2, 3/4, 5/6, 7/8} {
			\draw[line width=1.6pt] (s\i) to[bend left=60] (s\j);
		}
		
		% Level p=2:
		\draw[line width=0.8pt] (1.5, 0.3) to[bend left=60] (3.5, 0.3);
		\draw[line width=0.8pt] (5.5, 0.3) to[bend left=60] (7.5, 0.3);
		
		% Level p=3:
		\draw[line width=0.2pt] (2.5, 0.8) to[bend left=40] (6.5, 0.8);
		\node[draw=none, fill=none] at (4.5*\xsep, 1.3) {$p=N$};
		
		% External field arrow
		\draw[->, red, thick] (8.5, -0.2) -- (8.5, 0.6);
		\node[draw=none, fill=none, red] at (8.7, 0.2) {$h$};
	\end{tikzpicture}
	\caption{Schematic representation of the hierarchical structure of interactions in the DHM with $N=3$ levels and $L=8$ sites. The weakest interaction corresponds to the top-level $p=N$. A possible external field $h$ couples linearly to the spins $s_i$ at the sites $i \in \{1,\dots,8\}$.}
	\label{fig: DHM}
\end{figure}

In this section, we focus on a lattice system different from the 1D LRI model. The %underlying 
microscopic variables are still Ising spins $s_i \in \{ -1,+1\}$ along a one-dimensional chain, but the hierarchical model introduced by Dyson (DHM) in~\cite{dyson1969existence} is most naturally described in terms of block-spin variables. The hierarchical nature of the model is captured by a label $p \in \{0,1,\dots,N\}$ which we refer to as level. The number $N$ of levels is a positive integer. At the lowest level $p=0$, the block-spin variables are exactly the same as the original spins $\{s_i\}$, and we have $2^N \equiv L$ of them. At level $p=1$, we join adjacent variables into two-spin blocks, resulting in $2^N/2 = 2^{N-1}$ block-spins. At the higher levels, we join again two adjacent blocks to form bigger ones. In general, $r \in \{1,2,\dots, 2^{N-p}\}$ identifies a specific block, once $p$ has been fixed. Of course, the last level $p=N$ corresponds to having all the original Ising spins $\{s_i\}$ grouped into a single block variable. Fig.~\ref{fig: DHM} provides a visual summary of the above description. Mathematically, this whole procedure is encoded by defining the block-spin variables as the following sums
\begin{equation}\label{eq: definition of block spins}
	S_{p,r} = \sum_{i = 1+2^p(r-1)}^{2^p r} s_i.
\end{equation}
It is evident that the recursive property $S_{p-1,2r-1} + S_{p-1,2r} = S_{p,r}$, which enables us to pass from level $p-1$ to level $p$, must hold. Moreover, according to the description above, it is also true that each variable $S_{p,r_1}$ is equivalent to any other $S_{p,r_2}$ at the same level. This is related to the very large symmetry group of the DHM, which has order $2^{L-1}$ and is ultimately responsible for the abundance of exact results about it~\cite{meurice2007nonlinear}.

The idea behind the DHM is to emulate the long-range interactions of the form $|i-j|^{-(1+\sigma)} s_i s_j$ in the 1D LRI model. In order to obtain two-body interactions $s_i s_j$ it is sufficient to square the variables $S_{p,r}$ defined in~\eqref{eq: definition of block spins}. This generates interactions between any two spins $s_i$ and $s_j$ contained in the $r$-th block at level $p$. However, inter-block interactions are absent. To capture the decay of interactions with distance, the intensity of the coupling is suppressed as the block size increases with the level $p$, by multiplying $S_{p,r}^2$ by $2^{-p(1+\sigma)}$. In fact, the Hamiltonian of the DHM reads~\cite{dyson1969existence,meurice2007nonlinear}
\begin{equation}\label{eq: Hamiltonian of the  DHM}
	H_N = - \sum_{p=1}^N \frac{J}{2^{p(1+\sigma)}} \sum_{r=1}^{2^{N-p}} (S_{p,r})^2,
\end{equation}
with some constant $J$. The two main goals of this section -- i.e.~the comparison with the 1D LRI model and the renormalization of the DHM -- are achieved by rewriting the Hamiltonian~\eqref{eq: Hamiltonian of the  DHM} in terms of an ultrametric distance and in a recursive fashion, respectively.

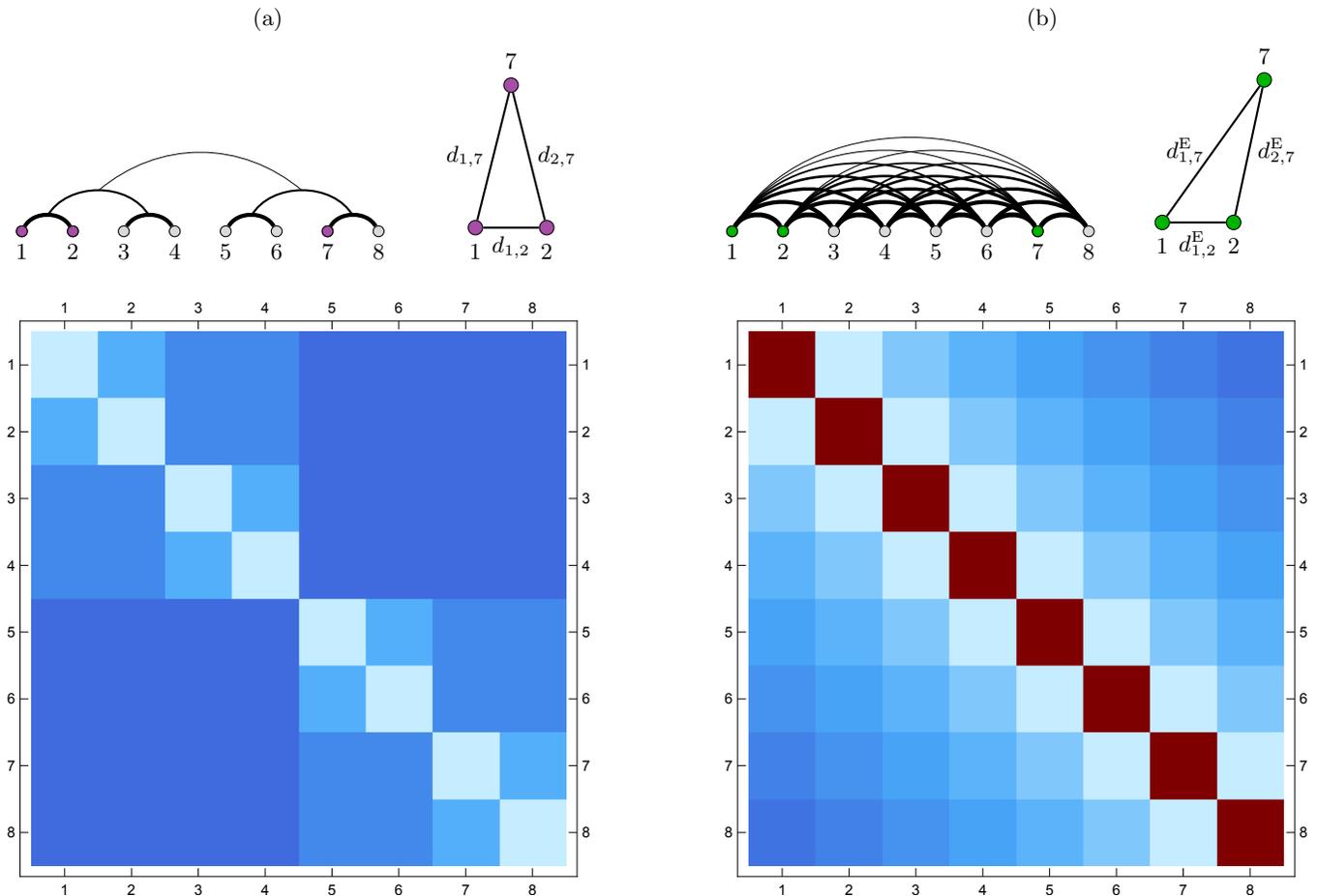
\begin{figure*}
    \centering
    {(a)} \hspace{10cm} {(b)}
    \smallskip\\
    \begin{tikzpicture}[scale=0.7, every node/.style={circle, draw, fill=gray!30, inner sep=1.5pt}]
		
		% Define spacing
		\def\xsep{1}
		\def\ybase{0}
		\def\yarcone{0.75}
		\def\yarctwo{1.8}
		
		% Draw sites
		\foreach \i in {1,...,8} {
			\node (s\i) at ({\i*\xsep}, \ybase) {};
			\node[draw=none, fill=none] at ({\i*\xsep}, \ybase - 0.4) {\i};
		}

        \foreach \x in {1,2,7} {\node[circle, draw, fill=violet!70, inner sep=1.5pt] at (\x,0) {};}
		
		% Level p=1:
		\foreach \i/\j in {1/2, 3/4, 5/6, 7/8} {
			\draw[line width=1.6pt] (s\i) to[bend left=60] (s\j);
		}
		
		% Level p=2:
		\draw[line width=0.8pt] (1.5, 0.3) to[bend left=60] (3.5, 0.3);
		\draw[line width=0.8pt] (5.5, 0.3) to[bend left=60] (7.5, 0.3);
		
		% Level p=3:
		\draw[line width=0.2pt] (2.5, 0.8) to[bend left=40] (6.5, 0.8);
	\end{tikzpicture}
    \hspace{0.4cm}
    % ===== Ultrametric triangle =====
\begin{tikzpicture}[scale=0.7]
  \node[circle,draw,fill=violet!70,inner sep=2pt,label=below:{$1$}] (A) at (0,0) {};
  \node[circle,draw,fill=violet!70,inner sep=2pt,label=below:{$2$}] (B) at (1.4,0) {};
  \node[circle,draw,fill=violet!70,inner sep=2pt,label=above:{$7$}] (C) at (0.7,2.8) {};
  \draw[thick] (A)--(B) node[midway,below] {$d_{1,2}$};
  \draw[thick] (A)--(C) node[midway,left]  {$d_{1,7}$};
  \draw[thick] (B)--(C) node[midway,right] {$d_{2,7}$};
\end{tikzpicture}
    \hfill
    \begin{tikzpicture}[scale=0.7, every node/.style={circle, draw, fill=gray!30, inner sep=1.5pt}]
		
		% Define spacing
		\def\xsep{1}
		\def\ybase{0}
		\def\yarcone{0.75}
		\def\yarctwo{1.8}
		
		% Draw sites
		\foreach \i in {1,...,8} {
			\node (s\i) at ({\i*\xsep}, \ybase) {};
			\node[draw=none, fill=none] at ({\i*\xsep}, \ybase - 0.4) {\i};
		}

        \foreach \x in {1,2,7} {\node[circle, draw, fill=green!70!black, inner sep=1.5pt] at (\x,0) {};}
		
		% |i-j|=1:
		\foreach \i/\j in {1/2,2/3,3/4,4/5,5/6,6/7,7/8} {
			\draw[line width=1.9pt] (s\i) to[bend left=60] (s\j);
		}
		
		% |i-j|=2:
        \foreach \i/\j in {1/3,2/4,3/5,4/6,5/7,6/8} {
			\draw[line width=1.6pt] (s\i) to[bend left=60] (s\j);
		}

        % |i-j|=3:
        \foreach \i/\j in {1/4,2/5,3/6,4/7,5/8} {
			\draw[line width=1.3pt] (s\i) to[bend left=60] (s\j);
		}

        % |i-j|=4:
        \foreach \i/\j in {1/5,2/6,3/7,4/8} {
			\draw[line width=1.0pt] (s\i) to[bend left=60] (s\j);
		}
        
        % |i-j|=5:
        \foreach \i/\j in {1/6,2/7,3/8} {
			\draw[line width=0.7pt] (s\i) to[bend left=60] (s\j);
		}

        % |i-j|=6:
        \foreach \i/\j in {1/7,2/8} {
			\draw[line width=0.4pt] (s\i) to[bend left=60] (s\j);
		}

        % |i-j|=7:
        \foreach \i/\j in {1/8} {
			\draw[line width=0.1pt] (s\i) to[bend left=60] (s\j);
		}
		
	\end{tikzpicture}
    \hspace{0.4cm}
    % ===== Euclidean triangle =====
\begin{tikzpicture}[scale=0.7]
  \node[circle,draw,fill=green!70!black,inner sep=2pt,label=below:{$1$}] (A) at (0,0) {};
  \node[circle,draw,fill=green!70!black,inner sep=2pt,label=below:{$2$}] (B) at (1.4,0) {};
  \node[circle,draw,fill=green!70!black,inner sep=2pt,label=above:{$7$}] (C) at (2.0,2.8) {};
  \draw[thick] (A)--(B) node[midway,below] {$d^{\rm E}_{1,2}$};
  \draw[thick] (B)--(C) node[midway,right] {$d^{\rm E}_{2,7}$};
  \draw[thick] (A)--(C) node[midway,left] {$d^{\rm E}_{1,7}$};
\end{tikzpicture}
    \bigskip\\
    \includegraphics[width=0.45\linewidth]{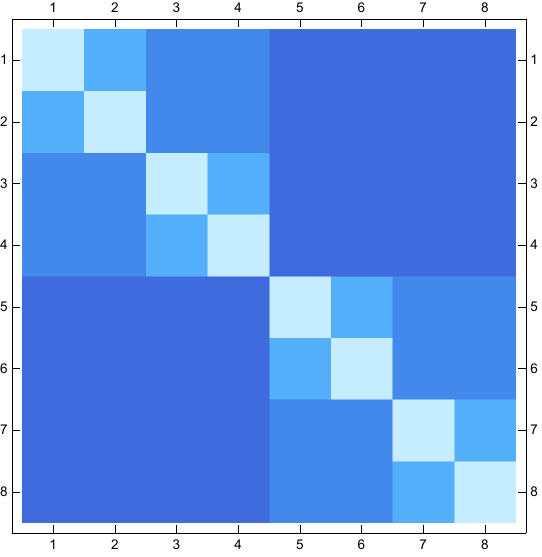}
    \hfill
    \includegraphics[width=0.45\linewidth]{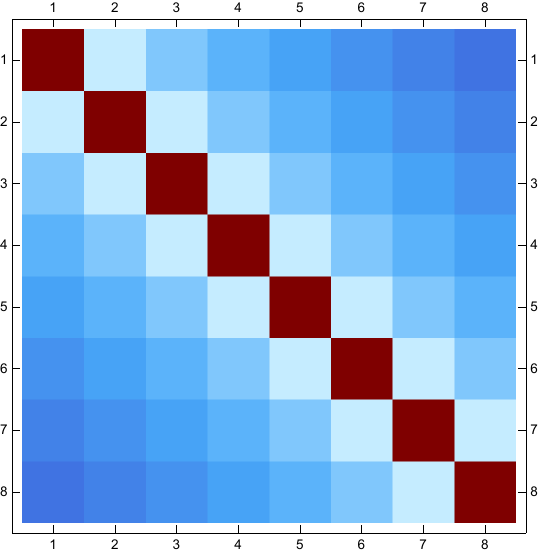}
    
    \caption{One-dimensional chains of $L=8$ Ising spins with either (a) hierarchical or (b) long-range interactions. 
    (a) The interaction matrix elements~\eqref{eq: matrix elements adjacency} of the hierarchical model exhibit ultrametric structure, since they depend on the distance $d_{ij}$, measuring the depth of the lowest common ancestor of two `leaves' (spins). Arranging distances as a triangle (shown for sites $1$, $2$ and $7$), the ultrametric inequality forces an isosceles configuration with a short base.
    (b) The intensity of matrix elements~\eqref{eq:LR_Hamiltonian} in the 1D LRI decays as a power law $J_{ij}\sim |i-j|^{-(1+\sigma)}$ of the Euclidean distance $d^{\rm E}_{ij} = |i-j|$. Contrary to (a), Euclidean distances form a generic scalene triangle, for which the usual triangle inequality holds. In these figures, lighter colors correspond to larger numerical values of the matrix elements, while the dark diagonal spots in (b) correspond to $J_{ii}=0$.}
    \label{fig:matrices}
\end{figure*}

Let us begin with the first task. In order to construct a fully-connected Ising-like Hamiltonian~\cite{capizzi2022spreading}
\begin{equation}\label{eq: Hamiltonian of DHM with ultrametric dist}
	H_N = - J \sum_{ij} A_{ij} s_i s_j, \quad i,j \in \{1,\dots, L=2^N\},
\end{equation}
we have to evaluate the cumulative interaction between spins $s_i$ and $s_j$ across all hierarchical levels. One starts from $p = d_{ij}$, the first level at which the sites $i$ and $j$ belong to the same block. The function $d_{ij}$ satisfies the property of a distance in an ultrametric space~\cite{agliari2017exact}, where the triangle inequality is replaced by the condition $d_{ij} \leq \max \{d_{ik}, d_{kj}\}$. As a result, the matrix elements $A_{ij}$ in Eq.~\eqref{eq: Hamiltonian of DHM with ultrametric dist} are
\begin{equation}\label{eq: matrix elements adjacency}
	A_{ij} = \sum_{p=d_{ij}}^{N} \frac{1}{2^{p \alpha} } = \frac{2^{-[d_{ij}-1]\alpha} - 2^{-N \alpha}}{2^\alpha -1},
\end{equation}
where we are using the notation $\alpha = 1+\sigma$ for brevity. It is worth noting that the $L\times L$ matrix $\mathds{A}$ with entries given by~\eqref{eq: matrix elements adjacency} possesses the block-hierarchical structure shown in Fig.~\ref{fig:matrices}(a). 

This contrasts with the 1D LRI model, cf.~Fig.~\ref{fig:matrices}(b), where the interactions~\eqref{eq:LR_Hamiltonian} depend directly on the Euclidean distance and no tree hierarchy is present. In the thermodynamic limit $N, L \to \infty$ the correction $2^{-N \alpha}$ in~\eqref{eq: matrix elements adjacency} becomes negligible and $A_{ij} \sim 2^{-d_{ij} \alpha}$. Comparing with~\eqref{eq:LR_Hamiltonian}, the correspondence $2^{d_{ij}} \approx |i-j|$ between hierarchical and Euclidean distance may be drawn~\cite{capizzi2022spreading}.

We remark in passing that the field of p-adic numbers, which can be represented as infinite paths in a rooted tree (in our case a 2-adic tree), plays an important role in the rigorous RG-based formulation of quantum field theory, see e.g.~\cite{abdesselam2013rigorous,abdesselam2018towards}.

Moreover, in~\eqref{eq: Hamiltonian of DHM with ultrametric dist} we include the constant term corresponding to the level $p=0$ of the microscopic spins ($S_{0,i} = s_i$) via the diagonal elements with $d_{ii} = 0$. In the matrix plot of Fig.~\ref{fig:matrices}(a), these are visualized as bright spots along the diagonal, as opposed to Fig.~\ref{fig:matrices}(b), where $J_{ii} = 0$ for the 1D LRI model. One can check that the inclusion of the level $p=0$ ensures the matrix $\mathbb{A}$ is positive definite, a property we will exploit in Sec.~\ref{sec3B}, where we also motivate the inclusion of this term -- originally absent from the Hamiltonian~\eqref{eq: Hamiltonian of the  DHM}.

Finally, we turn to the recursive form of the Hamiltonian~\eqref{eq: Hamiltonian of the  DHM}. In analogy with Ref.~\cite{angelini2014relations}, we write
\begin{align}\label{eq: recursivity of the Hamiltonian}
	H_{N} = H_{N-1}^{\rm left} + H_{N-1}^{\rm right} - \frac{J}{2^{N(1+\sigma)}} S_{N,1}^2,
\end{align}
which is just the expression of the possibility of joining two $(N-1)$-level models and adding interactions at the top level (in fact, $S_{N,1} = s_1+\dots+s_L$) in order to obtain an $N$-level DHM. More precisely, $H_{N-1}^{\rm left}$ involves the spins at sites $\{1,\dots, L/2\}$, while $H_{N-1}^{\rm right}$ refers to the sites $\{L/2+1,\dots,L\}$.

The self-similar nature of hierarchical tree structures can be used to derive exact RG equations in several different context, from critical phenomena~\cite{kim1977critical} to polymers~\cite{derrida1988polymers}, information transitions~\cite{gerbino2025measurement} and approximated ones for quantum phase transitions~\cite{monthus2015dyson}. Before proceeding to discuss the renormalization of the classical DHM, we remark that the model displays a phase transition for all values of $\alpha \in (1,2)$ -- or $\sigma \in (0,1)$ -- like the 1D LRI model~\cite{dyson1969existence}. The case $\sigma = 1$ is more delicate, because the 1D LRI model has a phase transition with BKT features~\cite{thouless1969long,frohlich1982phase}, while the DHM does not~\cite{dyson1969existence}. Nonetheless, a slight modification of the decay of the interaction in Eq.~\eqref{eq: Hamiltonian of the  DHM} -- namely replacing the factor $2^{-p(1+\sigma)} = 2^{-2p}$ by $2^{-2p} \log(p)$ -- \textit{does} yield a BKT-like transition and the Thouless effect~\cite{dyson1971ising}. We leave a more careful study of this variation of the DHM to future investigations.

\subsection{Real-space renormalization}
For later convenience, we add to the Hamiltonian $H_N$ an external field term of the form $- h S_{N,1}$, such that each individual spin $s_i$ interacts linearly with the field $h$. Even so, the recursive property~\eqref{eq: recursivity of the Hamiltonian} holds unchanged.

Hence, using~\eqref{eq: recursivity of the Hamiltonian} and taking into account the possibility of external fields, the partition function satisfies
\begin{equation}
	Z_N(\beta,h) = \sum_{\{s_i\}} e^{-\beta[H_{N-1}^{\rm left}(h) + H_{N-1}^{\rm right}(h)]} e^{A_N S_{N,1}^2}
\end{equation}
where $A_N \equiv 2^{-N(1+\sigma)} \beta J$. A Hubbard-Stratonovich transformation enables us to recast the last exponential as a Gaussian integral over an auxiliary variable $\phi$, so that
\begin{align}
	& Z_N(\beta,h) = \frac{1}{\sqrt{\pi}} \int_{-\infty}^{\infty} d\phi \, e^{-\phi^2} \times
	\nonumber\\
	& \times \sum_{\{s_i\}} e^{-\beta[H_{N-1}^{\rm left}(h) + H_{N-1}^{\rm right}(h)] + 2\sqrt{A_N} \phi S_{N,1}}.
\end{align}
Redefining the external field as $h \mapsto \hat{h}_N = h + 2\sqrt{A_N} \phi \beta^{-1}$, the partition function reduces to the one of a system with $N-1$ levels and an external field $\hat{h}_N$, i.e.
\begin{align}\label{eq: recursivity of partition function}
	& Z_N(\beta,h) = \frac{1}{\sqrt{\pi}} \int_{-\infty}^{\infty} d\phi \, e^{-\phi^2} Z_{N-1}(\beta, \hat{h}_N)^2.
\end{align}
In summary, we effectively removed the weakest link in Fig.~\ref{fig: DHM}, in order to decouple the left and right chains, by modifying the external field $h \mapsto \hat{h}_N(\phi)$ in the Hubbard-Stratonovich representation. 

Since the recursive equation~\eqref{eq: recursivity of partition function} is true at each level, $N$ can be replaced by $p \in \{1,\dots, N\}$. Moreover, reading~\eqref{eq: recursivity of partition function} from right to left, we can interpret it as a block-spin transformation in the RG sense~\cite{kadanoff1966scaling}, with length scale factor $\ell = 2$. Even after one RG step -- as we have anticipated -- the form of the Hamiltonian stays unchanged: Possible nonlocal interactions are not generated by the RG due to the hierarchical structure of the system. This already suggests a deep similarity with the LPA discussed in Sec.~\ref{sec2}. 

Introducing $\tilde{P}_{p}(\beta A_p^{-1/2} x/\sqrt{2}) = Z_{p-1}(\beta, x)^2$, one obtains
\begin{align}\label{eq: RGE dual spin distribution}
	& \tilde{P}_{p+1}(2^{\frac{1+\sigma}{2}} y) =\left[\frac{1}{\sqrt{\pi}} \int_{-\infty}^{\infty} d\varphi \, e^{-(\varphi-y)^2} \tilde{P}_{p}(\varphi) \right]^2,
\end{align}
after the change of variable from $\phi$ to $\varphi = \phi+y$, with $y =  \beta A_p^{-1/2} h/\sqrt{2}$. This RG equation can be taken as the starting point for the analysis of the fixed points of the DHM.

An alternative route to the RG procedure~\cite{kim1977critical,meurice2007nonlinear} is to introduce the following probability measure for a spin variable $s$
\begin{equation}\label{eq: UV measure}
	P_0(s) = \frac{1}{2}[\delta(s-1)+\delta(s+1)]
\end{equation}
so that the partition function reads
\begin{equation}
	Z_N(\beta,h) = \sum_{\{s_i\}} e^{-\beta H_N} = \int \left[ \prod_{i=1}^{L} P_0(s_i) d s_i \right]  e^{-\beta H_N},
\end{equation}
and to derive an effective RG equation for this quantity. In particular, the probability $P_p(s)$ after $p$ steps of renormalization evolves according to~\cite{kim1977critical,meurice2007nonlinear}
\begin{equation}\label{eq: RGE spin distribution}
	P_{p+1}( \sqrt{C} s) = 2^{\frac{1+\sigma}{2}} e^{- \beta C s^2} \int_{-\infty}^{\infty} dy \, P_p(s+y) P_p(s-y),
\end{equation}
where $C = 2^{1-\sigma}$.
In passing we note that the RG transformation involves a rescaling of the spin by a factor proportional to $\ell^{-\frac{1-\sigma}{2}}$, with $\ell = 2$. Comparing that to the conventional rescaling $\ell^{-\frac{d-2+\eta}{2}}$ in $d=1$, we obtain $\eta = 2-\sigma$, which is the same as the Sak relation~\cite{sak1973recursion} for long-range models, including the 1D LRI model.

Eq.~\eqref{eq: RGE spin distribution} contains the same information as~\eqref{eq: RGE dual spin distribution}, which now can be interpreted~\cite{kim1977critical} as the RG equation for the Hubbard-Stratonovich dual $\tilde{P}_p$ of the spin measure $P_p$:
\begin{equation}\label{eq: HS transform}
	\tilde{P}_p(\varphi) = \int_{-\infty}^{\infty} ds \, e^{-\beta J s^2 + 2\sqrt{\beta J} s \varphi} P_p(s).
\end{equation}
Indeed, as discussed further in the next section, $\tilde{P}(\phi)$ can be interpreted as the probability distribution of the variables $\phi_i$, conjugate to the spin variables $s_i$.

\subsection{A local potential field theory for the DHM}\label{sec3B}

As discussed earlier, the $p=0$ term $-J \sum_i s_i^2$ is added to the Hamiltonian of the DHM. This is achieved while keeping the partition function invariant, that is
\begin{align}
	Z_N &= \int  e^{-\beta H_N} \prod_{i=1}^{L} P(s_i) d s_i 
	\nonumber\\
	&= \int e^{\beta J \sum_{ij}  A_{ij} s_i s_j} \prod_{i=1}^{L} e^{-\beta J s_i^2} P(s_i) d s_i,
\end{align}
where the last line includes the $p=0$ level and therefore features the Hamiltonian~\eqref{eq: Hamiltonian of DHM with ultrametric dist} written in terms of the ultrametric matrix $A_{ij}$, which is positive definite. Due to the latter property, we can again use the Hubbard-Stratonovich transformation from the spins $\{s_i\}$ to the field variables $\{\phi_i\}$, obtaining
\begin{align}
	Z_N
	&\propto \int e^{- \sum_{ij}  \phi_i A^{-1}_{ij} \phi_j} \prod_{i=1}^{L} \tilde{P}(\phi_i) d \phi_i,
\end{align}
where $\tilde{P}$ is given by~\eqref{eq: HS transform}. Now, we separate off the diagonal contribution from the bilinear term
\begin{equation}
	A^{-1}_{ij} = K_{ij} + m^2 \delta_{ij},
\end{equation}
and we define a local potential
\begin{equation}
	V(\phi_i) = m^2 \phi_i^2 - \log \tilde{P}(\phi_i).
\end{equation}
Therefore, the partition function $Z_N = \int d^L \phi \, e^{-H^{\rm eff}_N[\phi]}$ can be written in terms of the effective Hamiltonian
\begin{equation}\label{eq: effective Hamiltonian DHM}
	H^{\rm eff}_N[\phi] = \sum_{i \neq j} \phi_i K_{ij} \phi_j + \sum_{i} V(\phi_i).
\end{equation}

In this form, we can notice that the DHM renormalizes in exactly the same way as the LPA~\eqref{eq: LPA ansatz}: The gradient term remains untouched, while the local potential term flows under RG as a consequence of Eq.~\eqref{eq: RGE dual spin distribution}. In the DHM this occurs by construction, while the absence of renormalization of the gradient term in the LPA in momentum-space RG is considered a truncation of the full theory. Hence, we suggest that the latter truncation in the field-theoretical framework corresponds to approximating the 1D LRI model on the lattice by the DHM (with the same $\sigma$).

In order to make the analogy even tighter, we now %demonstrate 
show that the gradient term in~\eqref{eq: effective Hamiltonian DHM} mimics the non-analytic momentum dependence $q^\sigma$ appearing in~\eqref{eq: LPA ansatz}. In fact, the eigenvalues of $\mathds{A} = (A_{ij})$ read
\begin{equation}
    \lambda_k = \sum_{j = 0}^{N-k} 2^{-j \sigma} = \frac{2^\sigma-2^{-(N-k)\sigma}}{2^\sigma-1}
\end{equation}
for $k \in \{0, \dots, N\}$. Their corresponding multiplicities are given by $2^{k-1}$ for $k>0$, with the largest eigenvalue $\lambda_0$ being non-degenerate. This expression is compatible with the spectrum of Ref.~\cite{agliari2017exact}, where the level $p=0$ was however not included in the coupling matrix. The eigenvectors $\boldsymbol{v}_k$, though known~\cite{agliari2017exact,capizzi2022spreading}, do not correspond to plane waves, as the DHM does not exhibit translational invariance. First, both $\boldsymbol{v}_0$ and $\boldsymbol{v}_1$ contain $2^N$ non-zero components. While $\boldsymbol{v}_0$ -- up to normalization factors -- is the constant vector $(1,1,\dots,1)$, the first half of $\boldsymbol{v}_1$ is filled with $-1$ and its second half with $+1$. Then, in general the $k$-th eigenspace is spanned by the mutually orthogonal eigenvectors $\boldsymbol{v}_k^{(m)}$, with $m \in \{1,\dots, 2^{k-1}\}$ and constructed as follows. The only nonzero elements of the vector $\boldsymbol{v}_k^{(1)}$ are the first $2^{N-k+1}$ ones: The first half of them are all equal to $-1$, the second half are all $+1$. Next, $\boldsymbol{v}_k^{(2)}$ contains the same elements, but shifted: The first $2^{N-k+1}$ ones are now zeros, but the second batch of $2^{N-k+1}$ elements are filled exactly as for $\boldsymbol{v}_k^{(1)}$. Iterating this procedure yields all the other eigenvectors.

\begin{figure}
    \centering
    \includegraphics[width=\linewidth]{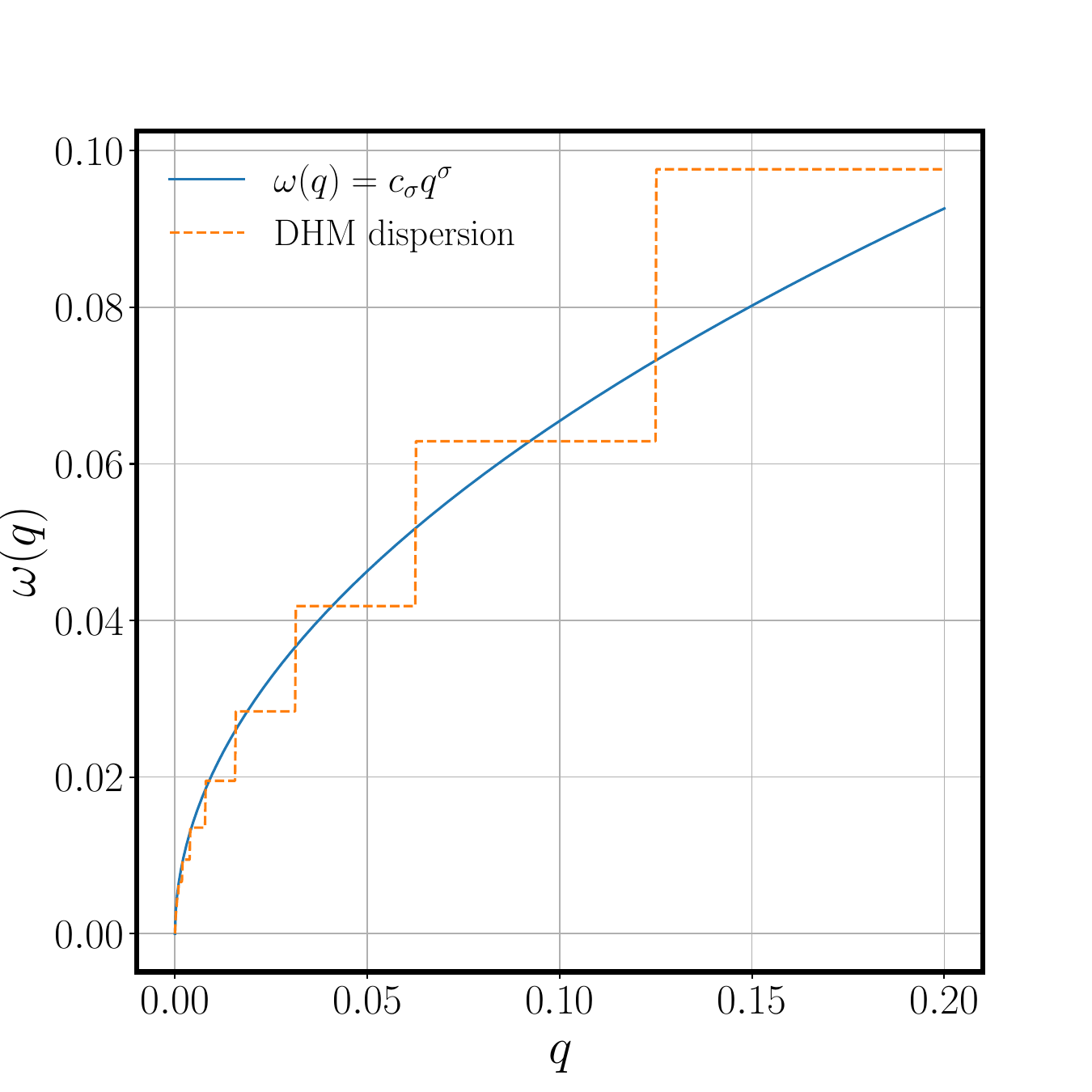}
    \caption{Comparison between the long-range dispersion $\omega(q) = c_\sigma q^\sigma$ and its hierarchical counterpart $\omega_k$ from Eq.~\eqref{eq:weddingcake}. For this plot $\sigma = 0.5$ and $N=20$ have been chosen.}
    \label{fig:weddingcake}
\end{figure}

Knowing the spectrum of the adjacency matrix $\mathbb{A}$, the eigenvalues of $\mathbb{K}$ are then given by $\lambda_k^{-1}-m^2$. Since we interpret the first term in the effective Hamiltonian $H^{\rm eff}$ as a gradient term, we select a value of $m^2$ such that the zero mode ($k=0$) has zero energy, i.e.~$m^2 = \lambda_0^{-1}$. We conclude that the spectrum of the gradient term is
\begin{align}
    \omega_k &= \lambda_k^{-1} - \lambda_0^{-1},
\end{align}
with the same degeneracy as the $\lambda_k$.
It is now useful to define the pseudo-momentum $q_k \equiv 2^{-(N-k)}$, with $0 < q_k \leq 1$, such that
\begin{equation}\label{eq:weddingcake}
    \omega_k = \omega(q_k) = c_\sigma \frac{q_k^\sigma}{1-(q_k/2)^\sigma},
\end{equation}
up to terms that vanish in the thermodynamic limit $N\to \infty$. The prefactor is $c_\sigma = 2^{-\sigma}(1-2^{-\sigma})$. In a continuous representation we can write the dispersion
\begin{equation}\label{eq:pseudo-dispersion}
    \omega(q) = \sum_{k = 0}^{N-1} \omega(q_k) \theta(q-q_k)\theta(q_{k+1}-q),
\end{equation}
where $q \in (0,1]$. In the thermodynamic limit, for sufficiently small values of $q_k$, we observe that $\omega(q) \sim c_\sigma q^{\sigma}$, as visualized in Fig.~\ref{fig:weddingcake}. 
Hence, we have shown that the hierarchical model produces a `wedding cake' version of the dispersion appearing in~\eqref{eq: gradient term}. Using the symbol $\lfloor -\Delta \rfloor^{\sigma/2}$ to denote a hierarchical Laplacian whose pseudo-momentum space representation is given by $\omega(q)$ in Eq.~\eqref{eq:pseudo-dispersion}, we compactly write the effective theory as
 \begin{equation}\label{eq: effective field theory of DHM}
	H^{\rm eff}[\phi] = \int_{-\infty}^{\infty} dx \, \left\{ \phi(x) \lfloor -\Delta \rfloor^{\sigma/2} \phi(x) + V(\phi(x)) \right\},
\end{equation}
after considering a suitable continuum limit. The analogy with~\eqref{eq: LPA ansatz} is now complete. 

Nevertheless, we expect that the hierarchical gradient term alters the specific form of the fixed point theory and also the values of the critical exponents. However, in the following discussion we are going to find out that the values of the exponent $\nu$ for the LPA and the DHM are numerically very close; in fact, they are almost equal within our current choice of the regulator, see Eq.\,\eqref{eq: Litim regulator}.

\subsection{Exponent $\nu$ and comparison with the LPA}\label{sec: DHM vs LPA}

In this part we review very briefly the procedure outlined in~\cite{kim1977critical} to obtain the critical exponents of the DHM. A convenient way of studying the RG equation~\eqref{eq: RGE dual spin distribution} starting from the initial condition $\tilde{P}_0(\phi)$ -- obtained from~\eqref{eq: UV measure} and~\eqref{eq: HS transform}, and dependent on the temperature $\beta^{-1}$ -- is to expand the dual spin measure $\tilde{P}_p(\phi)$ as a series of Hermite polynomials $H_k(a \phi)$, where $a = [1-2^{-(1+\sigma)}]^{1/2}$. This choice is prompted by the special property
\begin{equation}
	\int_{-\infty}^{\infty} \frac{d\xi}{\sqrt{\pi}} \, e^{-(\xi-\phi)^2} H_k(a \xi) = (1-a^2)^{\frac{k}{2}} H_k\left(\frac{a \, \phi}{\sqrt{1-a^2}}\right).
\end{equation}
Therefore, at each step $p$ one can write
\begin{equation}\label{eq: Hermite series}
	\tilde{P}_p(\phi) =  B_0^{(p)} + B_0^{(p)} \sum_{k=1}^{\infty} B_k^{(p)} 2^{k(1+\sigma)} H_{2k}(a \phi).
\end{equation}
Projecting the RG equation~\eqref{eq: RGE dual spin distribution} onto the coefficients $B_k^{(p)}$ of this expansion yields the recursion relations~\cite{kim1977critical}
\begin{subequations}
\begin{align}
	&B_0^{(p+1)} = \mu \, B_0^{(p)} B_0^{(p)},
	\\
	& 2^{k(1+\sigma)} \mu B_k^{(p+1)} = 2 B_k^{(p)} + \sum_{k',k''=1}^{\infty} T_{k k' k''} B_{k'}^{(p)} B_{k''}^{(p)}
\end{align}
\end{subequations}
for all $k\geq 1$, where we have defined
\begin{equation}\label{eq: defn of mu}
	\mu = 1+\sum_{k=1}^\infty 2^{2k} (2k)! \, B_k^{(p)} B_k^{(p)}
\end{equation}
and the combinatorial tensor
\begin{equation}
	T_{k k' k''} = \begin{cases}
		2^l l! \binom{2k'}{l} \binom{2k''}{l}, \quad & \text{if $|k'-k''|\leq k \leq k'+k''$,}
		\\
		0, \quad & \text{otherwise},
	\end{cases}
\end{equation}
with $l = -k + k'+k''$. When the temperature is tuned to its critical value $\beta = \beta_c$, entering the initial condition $\tilde{P}_0$ and thus the coefficients $\{B_k^{(p=0)}\}$, one is able to reach a fixed point characterized by $\{B_k^{*}\}$ and $\mu^*$, which is the obvious notation for~\eqref{eq: defn of mu} with $B_k^{(p)} \to B_k^*$. The fixed point can be found with very high accuracy and with a truncated series~\eqref{eq: Hermite series}, which typically need not include much more than $10$ terms~\cite{kim1977critical}. The rigorous construction of this \textit{interacting} fixed point, along with further aspects of the RG flow of Dyson-type hierarchical models, is explored in~\cite{bleher1973investigation,bleher1975critical,bleher2010critical}.

Once the fixed point is known, it is possible to linearize around it. One defines $\delta B_k^{(p)} = B_k^{(p)}-B_k^*$ and finds
\begin{equation}
	\delta B_k^{(p+1)} = \sum_{k'=1}^\infty V_{k k'} \, \delta B_k^{(p)},
\end{equation}
where the stability matrix is given by
\begin{equation}
	V_{k k'} = \frac{2}{\mu^*} \left( \frac{\delta_{k k'} + \sum_{k''=1}^{\infty} T_{k k' k''} B_{k''}^*}{2^{k(1+\sigma)}} - B_k^* B_{k'}^* 2^{2k'} (2k')! \right).
\end{equation}
The critical exponent $\nu$ is obtained from the largest eigenvalue $\lambda_1$ of the stability matrix,
\begin{equation}\label{eq: thermal eigenvalue DHM}
	\nu^{-1} = \log \lambda_1 / \log 2,
\end{equation}
under the requirement that there is only one unstable direction, and therefore all the other eigenvalues with $m\neq 1$ obey $|y_m| < 1$. 

We have carried out the calculation of the exponent $\nu$ for the DHM, and our results are shown as blue crosses in Fig.~\ref{fig:exponentnu}. One sees that $\nu^{-1} \to 1/2$ for $\sigma \to 1/2$ and $\nu^{-1} \to 0$ for $\sigma \to 1^{-}$,  as for the LPA results. Near $\sigma=1$ one has for the DHM results $\nu^{-1} \approx {\cal C} (1-\sigma)^\gamma$ with ${\cal C} \approx 4.0$ and $\gamma \approx 0.9$, again as for the LPA results.  

\begin{figure}[h]
	\centering
	\includegraphics[width=1.0\linewidth]{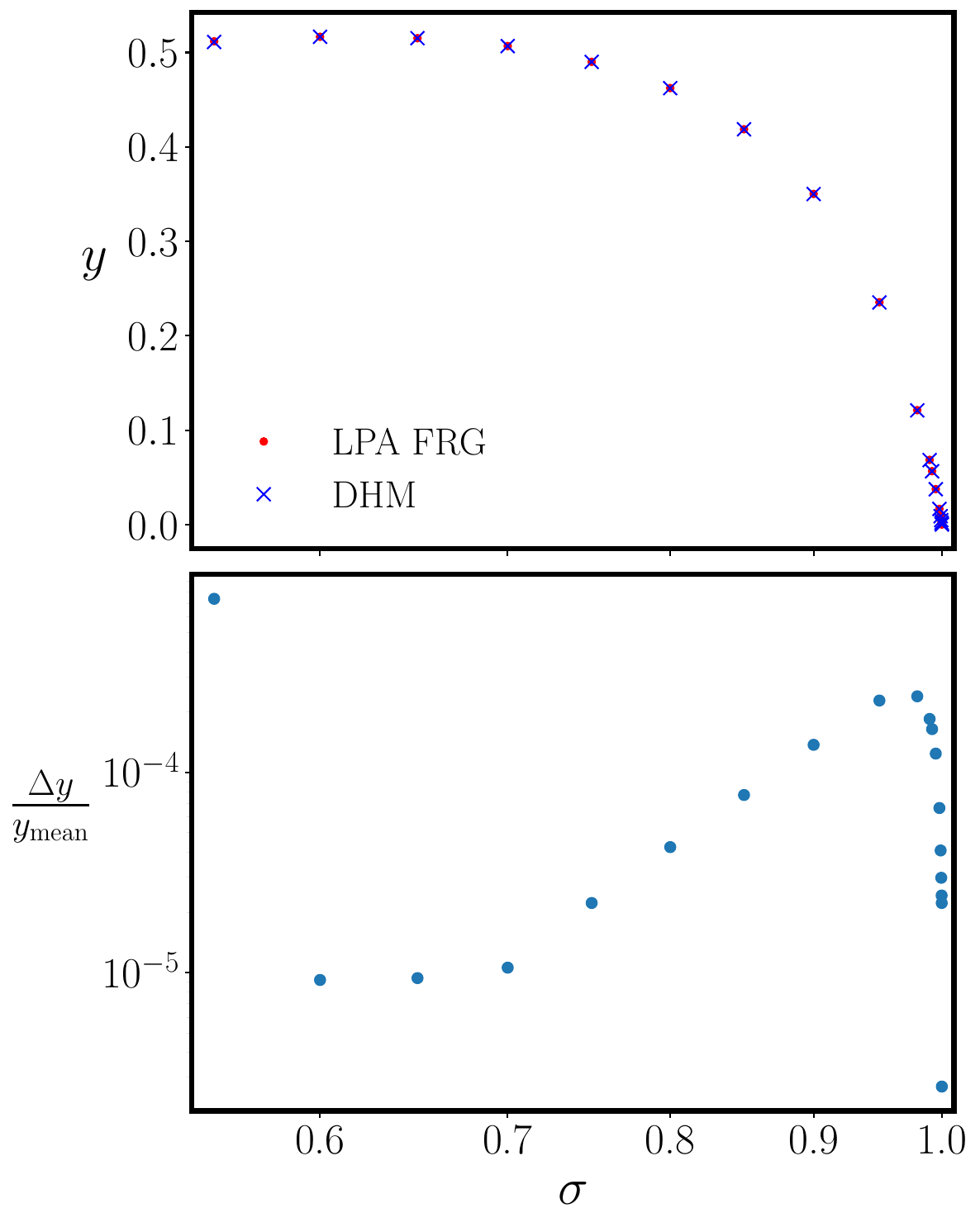}
	\caption{In the upper panel we show the thermal eigenvalue $y = \nu^{-1}$ obtained as solution of~\eqref{eq: eigenperturbation equation} in the LPA and~\eqref{eq: thermal eigenvalue DHM} for the DHM. The points overlap almost exactly, as in Fig.~\ref{fig:exponentnu}. Their relative difference, where $\Delta y = y_{\rm LPA}-y_{\rm DHM}$ and $y_{\rm mean} = (y_{\rm LPA}+y_{\rm DHM})/2$, is shown in the lower panel. The error is peaked in the region of large $\sigma$, but it stays below $10^{-3}$ throughout the interval $1/2<\sigma<1$.}
	\label{fig:differencelpahm}
\end{figure}

%Surprisingly, 
Remarkably, the values of $\nu^{-1}(\sigma)$ overlap \textit{almost perfectly} with the corresponding ones obtained via the FRG at the LPA level, as described in Sec.~\ref{sec2}. Their relative difference is always below $10^{-3}$ and actually mostly below $10^{-4}$ in the entire $\sigma$ range, see Fig.~\ref{fig:differencelpahm}.

The similarity between these two values comes as a surprise if one considers that the forms of the gradient terms in Eqs.~\eqref{eq: LPA ansatz} and~\eqref{eq: effective field theory of DHM} is different. In fact, the similarity only occurs for the present choice of the regulator~\eqref{eq: Litim regulator}, which is a long-range generalization of the optimized Litim regulator~\cite{litim2000optimization,litim2001optimized}. It was already noted in~\cite{litim2007towards} that the LPA of a short-range scalar theory in three dimensions reproduces, up to small numerical values, the exponent of a hierarchical model with rescaling factor $\ell^{1/3}$, provided that the optimized regulator~\cite{litim2000optimization,litim2001optimized} is used. Here, we demonstrate that this property extends over the entire $\sigma$ range and in $d=1$.

At an even earlier stage, it was discovered by Felder in~\cite{felder1987renormalization} that the $\ell \to 1$ limit of the RG recusion relations of the DHM is mathematically equal to the LPA in the exact RG formulation of Wilson and Polchinski. In turn, the latter is equivalent~\cite{morris2005equivalence} via a change of variables to the LPA in the Wetterich formulation -- the one that we have used in Sec.~\ref{sec2} -- upon choosing the optimized regulator proposed by Litim~\cite{litim2000optimization,litim2001optimized}.

Due to the fact that variations in the scale factor $\ell$ may change the universality class of the model, the natural formulation with blocks of size $\ell = 2$ in the DHM is \textit{not} mathematically equivalent to the LPA. However, the effect of such variation of $\ell$ in the critical exponents is minimal~\cite{meurice2007nonlinear}. This underlies the discrepancy observed in Fig.~\ref{fig:differencelpahm}.

\begin{figure}[h]
    \centering
    \includegraphics[width=1.0\linewidth]{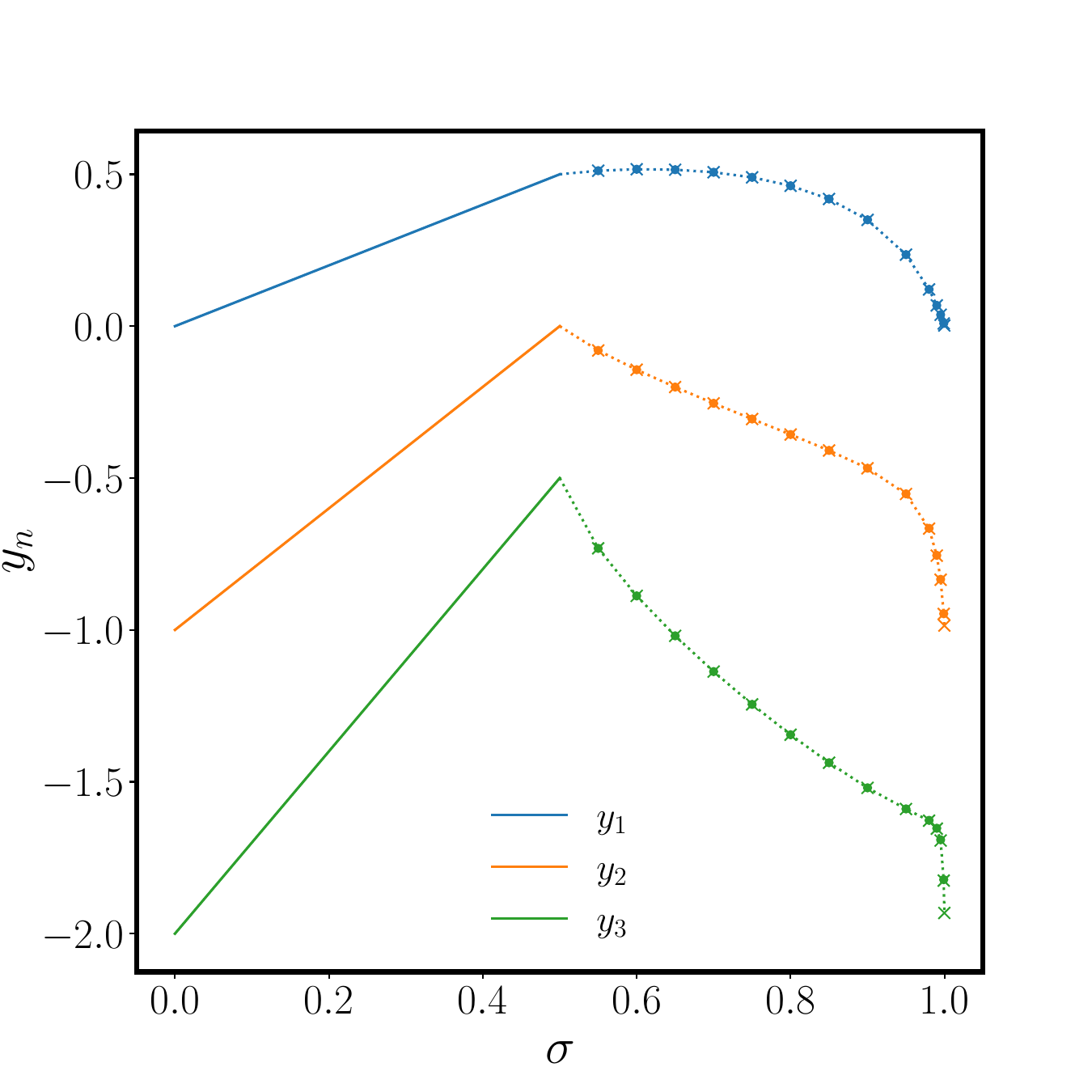}
    \caption{First three RG eigenvalues $y_1,y_2,y_3$ computed by means of the FRG of Sec.~\ref{sec2} (filled circles) or the RG for the DHM of Sec.~\ref{sec: DHM vs LPA} (crosses). We find close agreement between the latter methods around the interacting fixed-point, while the eigenvalues are exact (solid lines) for the Gaussian region $0<\sigma<1/2$.}
    \label{fig:eigenvalues}
\end{figure}

For completeness, we also report in Fig.~\ref{fig:eigenvalues} a comprehensive picture of the first few RG eigenvalues, where $y_1 = \nu^{-1}$ and $\omega = - y_2$ is the so-called `correction-to-scaling' exponent. In the region $0<\sigma\leq 1/2$ the Gaussian fixed point dictates the universal properties of the system and the eigenvalues are given by $y_n = n(\sigma-1)+1$~\cite{kim1977critical}. As we cross $\sigma=1/2$ towards larger values, as in Fig.~\ref{fig:differencelpahm}, the Gaussian fixed point exchanges stability with the interacting one, whose RG eigenvalues are computed via FRG (filled circles) or the method of~\cite{kim1977critical} (crosses). Not only the thermal eigenvalue $y_1$, but also the eigenvalues $y_{n>1}$ corresponding to irrelevant perturbations exhibit the LPA-DHM correspondence described previously. The discrepancy between the two methods is quantitatively comparable to that for $y_1$.

An interesting open question concerns the critical exponents of the theory~\eqref{eq: effective field theory of DHM} studied by means of the FRG with an appropriate generalization of the regulator~\eqref{eq: Litim regulator} where $q^\sigma$ would be replaced by $\omega(q)$. In that case, on the basis of the previous discussion, we expect to find exactly the same critical exponents as those of the DHM.

\section{Comparison with other methods}\label{sec4}

In this section, we compare our numerical estimates with known analytical and numerical results for the Ising model. For the sake of the following comparisons, the LPA and DHM are effectively equivalent, and we will refer mostly to the LPA of Sec.~\ref{sec2}.

In Ref.~\cite{fisher1972critical} the second-order $\epsilon$-expansion around the mean-field threshold $d=2\sigma$ was reported.
\begin{equation}\label{eq: epsilon expansion}
	\nu^{-1} = \frac{\sigma}{\gamma} = \sigma - \frac{\epsilon}{3} - \mathcal{A}(\sigma) \frac{\epsilon^2}{9} + O(\epsilon^3),
\end{equation}
where $\epsilon = 2\sigma-d = 2\sigma-1$, $\mathcal{A}(\sigma) = \psi(1)-2\psi(\sigma/2)+\psi(\sigma)$, and $\psi(\cdot)$ is the digamma function. In Fig.~\ref{fig:exponentnu} Eq.~\eqref{eq: epsilon expansion} is shown as the dotted black line and closely follows the LPA estimates in the region $\sigma \in (0.5, 0.55)$, while departing for $\sigma \gtrsim 0.6$.

A more thorough weak-coupling expansion was carried out for long-range models in~\cite{benedetti2020long,benedetti2025addendum}. This work provides three-loop estimates of the critical exponents in the $\epsilon$-expansion. In order to obtain accurate results, one additionally needs to perform a Padé-Borel resummation of the three-loop series. The results are plotted in Fig.~\ref{fig:exponentnu} as pink stars. At $\sigma = 0.6$, these higher-order corrections realign the perturbative results with the LPA, while at $\sigma = 0.7$ and $0.8$, the exponent $\nu^{-1}$ becomes compatible with \cite{tomita2008monte} and larger than the LPA one.

In the strong-coupling limit $\sigma \to 1$ the LPA reproduces the expected divergence of the correlation length exponent, but cannot capture the exact slope, which was obtained in the perturbative RG approach of Kosterlitz~\cite{kosterlitz1976phase}. Indeed, the alternating domain walls are appropriate weakly coupled degrees of freedom in the region close to $\sigma = 1$, see the end of Sec.~\ref{sec1} as well as Ref.~\cite{anderson1971some}. Within the domain wall formulation, one obtains $\eta(\sigma) = 2-\sigma$ and
\begin{equation}
	\nu(\sigma)^{-1} = \sqrt{2(1-\sigma)} + O(1-\sigma),
    \label{kosterlitz_domain_wall}
\end{equation}
which is the solid line of Fig.~\ref{fig:exponentnu}. Recently, Ref.~\cite{benedetti2025one} improved the result in Eq.~\eqref{kosterlitz_domain_wall}, giving for the scaling dimension of the dual of the $\phi^2$ field $\Delta_\epsilon=1-\sqrt{2 \delta}+\delta/4$, with $\delta=1-\sigma$. Using the relation between $\nu$ and $\Delta_\epsilon$ with $d=1$ in the results of~\cite{benedetti2025one} then gives
\begin{equation}
	\nu(\sigma)^{-1} = \sqrt{2(1-\sigma)} - \frac{1-\sigma}{4} + O((1-\sigma)^{3/2}).
    \label{cft_estimate}
\end{equation}
While the correction of~\cite{benedetti2025one} moves the curve of $\nu^{-1}$ (the dashed line in Fig.~\ref{fig:exponentnu}) closer to the LPA estimate, the slope of the two curves are still different at $\sigma \to 1^-$. One reason for the slope discrepancy may be the missing short-range term $q^{2}$ in the our LPA ansatz~\eqref{eq: LPA ansatz}. At larger dimensions, which also include quantum 1D LRI model, the FRG reproduces correctly both the weak and strong coupling limits close to $\sigma = \sigma_{\rm mf}$ and $\sigma = \sigma_*$, see Fig.\,(3a) of~\cite{defenu2017criticality}. As anticipated in the Introduction, we see that the topological effects at $\sigma \lesssim \sigma^* = 1$ become particularly strong, so that both the LPA~\eqref{eq: LPA ansatz} and the hierarchical approximation become unsatisfactory. The improvement of our theoretical description in this sense -- as well as the possibility of dealing with the field theory of~\cite{benedetti2025one,benedetti2025strong} within FRG -- is deferred to future work.

Consistently with the previous analysis, the comparison between the MC estimate of the critical exponents and the LPA results strongly depends on the $\sigma$ region, see Fig.~\ref{fig:exponentnu}. The MC results of Ref.~\cite{uzelac2001critical} (orange squares) approach the estimate in Eq.~\eqref{cft_estimate} for $\sigma \gtrsim 0.8$, while for $\sigma \lesssim 0.8$ they become compatible with the LPA prediction (small red circles). Furthermore, we have included a comparison with Ref.~\cite{tomita2008monte}, whose MC points (empty blue circles) are systematically higher than those of Ref.~\cite{uzelac2001critical}, but remain overall consistent with the previous analysis due to the larger error bars. The comparison with~\cite{uzelac2001critical} and~\cite{tomita2008monte} shows %demonstrates 
that the LPA yields accurate results in a large $\sigma$ interval, except in the %immediate 
vicinity of $\sigma = 1$.

A mapping was conjectured between the critical exponents value of the long-range model and those of the corresponding local model in an effective dimension $D_{\rm eff}$~\cite{banos2012correspondence}. This mapping has been shown to be remarkably %successful 
useful in the past~\cite{angelini2014relations,defenu2015fixed} and, although not supported beyond the LPA level~\cite{defenu2015fixed,behan2017scaling}, it produces accurate estimates, within 5\%, when applied to the exact values in $d=2$~\cite{solfanelli2024universality}. In the current $d=1$ case, one finds
\begin{align}\label{eq:effdim}
    \nu_{\rm LR}(\sigma) = D_{\rm eff} \nu_{\rm SR}(D_{\rm eff}),
\end{align}
with $\sigma = (2-\eta_{\rm SR}(D_{\rm eff}))/D_{\rm eff}$.
Then, we can take $D_{\rm eff}\in\{2,3,4\}$, $\eta_{\rm SR}(D_{\rm eff}) \in \{1/4,\eta_{\rm CB}, 0\}$, and $\nu_{\rm SR}(D_{\rm eff}) \in \{1, \nu_{\rm CB}, 1/2\}$, where the three-dimensional exponents $\eta_{\rm CB}$ and $\nu_{\rm CB}$ are the recent conformal bootstrap values taken from~\cite{chang2025bootstrapping}. The corresponding points are shown as green triangles in Fig.~\ref{fig:exponentnu}. Overall, the results of the effective dimension approach overshoot the MC results by Ref.~\cite{uzelac2001critical}. This is most likely the result of the approximate nature of the effective dimension approach.

In summary, the LPA study produces a critical exponent $\nu^{-1}(\sigma)$ well in agreement with the perturbative curve as $\sigma \to \sigma_{\rm mf} = 1/2$, while it remains consistently below the strong coupling expansion as $\sigma \to \sigma_*=1$. Based on the comparison with MC data and the effective dimension mapping, the accuracy of the LPA estimates is accurate up to $\sigma \approx 0.8$. Moreover, since the perturbative expansions and (almost) all Monte Carlo data points lie above our LPA/DHM curve, we %propose 
may conclude that the exponent $\nu^{-1}$ of the DHM %represents 
appears to be a lower bound for the exponent $\nu^{-1}$ of the 1D LRI model, in qualitative agreement with the general strategy of using the DHM to prove bounds for the 1D LRI.

\section{Conclusions}\label{sec5}
 
In this work, we analyzed the critical behavior of the one-dimensional long-range Ising (1D LRI) model by employing two complementary non-perturbative methods: the functional renormalization group (FRG) in the local potential approximation (LPA), and real-space renormalization of Dyson’s hierarchical model (DHM). Despite their conceptual differences — the LPA being a momentum-space continuum approach and the DHM a discrete, non-translationally invariant lattice model — we have shown that their predictions for the critical exponent $\nu$ are nearly indistinguishable across the entire long-range interacting regime $1/2 < \sigma < 1$.

We have argued that this remarkable agreement arises from a deeper structural analogy between the two formulations. In particular, both frameworks retain a fixed gradient term and renormalize only the local potential, with the DHM providing a constructive realization of this truncation. Furthermore, the use of the optimized Litim regulator~\eqref{eq: Litim regulator} in the LPA enhances this correspondence by mirroring the features of the hierarchical RG transformation.

Comparisons with perturbative expansions near $\sigma = 1/2$ and recent conformal field theory (CFT) results near $\sigma = 1$ provide a consistent validation of the LPA and DHM approximations, while also delineating their limitations. Our analysis underscores the difficulty of accurately capturing BKT-like transitions and topological phenomena as $\sigma$ approaches 1, where both approximations begin to falter (despite having the correct feature $\nu^{-1}$ for $\sigma \to 1^-$), primarily due to their neglect of short-range operators and topological defects. Nevertheless, Monte Carlo simulations indicate that the LPA retains quantitative reliability across a broad intermediate regime, extending up to $\sigma \approx 0.8$. Taken together, these findings yield a well-defined curve for the critical exponent $\nu$ in the classical Ising model in $d=1$. For $\sigma \lesssim 0.8$, the Ising curve closely follows the LPA prediction, while in the range $0.8 \lesssim \sigma \lesssim 1$, it is rather close to the CFT estimate given in Eq.~\eqref{cft_estimate} and consistently above LPA results.

Looking forward, several interesting directions emerge. %On the theoretical side, 
First, it would be desirable to improve upon the LPA by incorporating the effects of short-range analytic terms and testing the robustness of the observed equivalence beyond leading-order truncations. Additionally, adapting the FRG framework to study the effective DHM field theory -- with its hierarchical Laplacian dispersion $\omega(q)$ -- may allow for a direct and controlled comparison with numerical RG flows. From a broader perspective, the methods and analogies presented here may inform the treatment of more complex systems exhibiting long-range interactions, including quantum spin chains and out-of-equilibrium models~\cite{defenu2024out, molignini2024unbounded}.

Finally, the close correspondence between the DHM and the LPA offers not only computational advantages but also conceptual clarity in the understanding of long-range critical phenomena, suggesting that hierarchical models may serve as effective surrogates for more realistic systems in suitable regimes.

\noindent \textit{Note added.} During the completion of this work, we became aware of a related preprint~\cite{artun2025ferromagnetic}, analyzing the 1D LRI model via real-space RG, and reporting finite-temperature phases for the ferromagnetic and spin-glass regimes but no ordered phase in the antiferromagnetic case.
 
\begin{acknowledgments}
GG acknowledges the support of the MSCA Grant 101152898 (DREAMS). This research was funded by the Swiss National Science Foundation (SNSF) grant numbers 200021--207537 and 200021--236722, by the Deutsche Forschungsgemeinschaft (DFG, German Research Foundation) under Germany's Excellence Strategy EXC2181/1-390900948 (the Heidelberg STRUCTURES Excellence Cluster) and by the European Union under GA No. 101077500–QLRNet. Partial support by grant NSF PHY-230935 to the Kavli Institute for
Theoretical Physics (KITP) is also acknowledged.
	
\end{acknowledgments}

%%%%%%%%%%%%%%%%%%%%%%%%%%%%%%%%%%%%%%%%%%%%%%%%%%%%%%%%%%%%%

\appendix
\begingroup
\allowdisplaybreaks

%\bibliographystyle{apsrev4-2}
%\bibliography{ref_lib}


\end{document}